\documentclass[conference]{IEEEtran}
\usepackage[utf8]{inputenc}
\usepackage{amsmath,amssymb}
\usepackage{graphicx}
\usepackage{url}
\usepackage{lipsum}  % for placeholder text
\usepackage{balance}
\begin{document}
	
	\title{Pico-Cloud: Cloud Infrastructure for Tiny Edge Devices}
%	\title{Pico-Cloud: Local Tiny Cloud}
	\author{
		\IEEEauthorblockN{Mordechai Guri}
		\IEEEauthorblockA{Ben-Gurion University of the Negev, Israel\\Faculty of Computer and Information Science \\
			Email: gurim@post.bgu.ac.il}
	}
	
	\maketitle
\begin{abstract}
	This paper introduces the Pico-Cloud, a micro-edge cloud architecture built on ultra-minimal hardware platforms such as the Raspberry Pi Zero and comparable single-board computers. The Pico-Cloud delivers container-based virtualization, service discovery, and lightweight orchestration directly at the device layer, enabling local operation with low latency and low power consumption without reliance on centralized data centers. We present its architectural model, outline representative use cases including rural connectivity, educational clusters, and edge AI inference, and analyze design challenges in computation, networking, storage, and power management. The results highlight Pico-Clouds as a cost-effective, decentralized, and sustainable platform for lightweight distributed workloads at the network edge.
\end{abstract}

	
\section{Introduction}	
The Pico-Cloud is a miniature cloud infrastructure composed of low-cost, low-power computing devices such as the Raspberry Pi Zero and other ultra-compact single-board computers. It applies core cloud principles—including virtualization, containerization, and on-demand services—to a drastically reduced hardware footprint. In essence, the Pico-Cloud replicates many capabilities of public cloud platforms, enabling multiple isolated applications with integrated networking and storage, but within the constraints of small, resource-limited devices. This design supports cloud-native experimentation and provides localized services in environments where traditional data centers are impractical, such as remote locations, mobile deployments, or small-scale on-premises systems~\cite{Smith2016}.

From an architectural perspective, the Pico-Cloud emphasizes simplicity and decentralization. Each node operates on a bare-metal or lightweight operating system, prioritizing essential functionality while avoiding the complexity typically associated with large-scale platforms. The primary goal is to deliver reliable, scalable, cloud-like experiences using as few as one or two minimal nodes. Unlike centralized data centers, Pico-Cloud deployments are situated directly at the network edge or on-premises, close to data sources and end-users. This proximity leverages edge computing benefits such as reduced latency and improved data locality, while retaining cloud-oriented flexibility, including container deployment and lightweight orchestration~\cite{NodeWeaver2023}.

Relative to traditional cloud infrastructures and standard edge solutions, the Pico-Cloud represents the most minimal point on the distributed computing spectrum. Conventional cloud systems rely on powerful, centrally managed servers with high-bandwidth connectivity, while edge deployments typically depend on relatively strong, specialized, and costly hardware. In contrast, Pico-Cloud implementations rely on inexpensive, widely available hobbyist platforms, deliberately trading computational capacity for cost-effectiveness, portability, and autonomy. This minimalist approach fosters decentralization and user empowerment, enabling individuals and small communities to establish their own small-scale “cloudlets” tailored to local needs. Ultimately, the Pico-Cloud achieves cost efficiency, environmental sustainability, and simplified deployment by substantially reducing hardware complexity~\cite{Satyanarayanan2017,Bonomi2012}.

\section{Operational Domains and Use Case Scenarios}
Pico-Cloud infrastructures address a broad range of practical domains where conventional cloud or edge solutions may be infeasible, overpowered, or economically inaccessible. The use cases span multiple sectors and scenarios, demonstrating the versatility and practicality of Pico-Cloud technology.

%\begin{table*}[htbp]
%	\centering
%	\begin{tabular}{|l|p{12cm}|}
%		\hline
%		\textbf{Use Case} & \textbf{Short Description} \\
%		\hline
%		Edge AI Inference & Executing lightweight AI inference locally on-device to support latency-sensitive and disconnected applications. \\
%		\hline
%		Rural Connectivity & Enabling local services such as digital libraries, communication platforms, and public data in regions with poor internet access. \\
%		\hline
%		Educational Labs & Supporting accessible, affordable training in distributed systems and DevOps for schools and universities. \\
%		\hline
%		Disaster Recovery & Restoring essential IT and communications infrastructure post-disaster using power-efficient, portable cloud nodes. \\
%		\hline
%		Local Private Cloud & Providing privacy-preserving platforms for home or organizational file storage, web applications, or IoT control. \\
%		\hline
%		Innovation in Underserved Areas & Serving as infrastructure for grassroots IT innovations and low-cost civic or commercial deployments. \\
%		\hline
%		IoT Edge Gateway & Collecting, aggregating, and pre-processing data at the edge to reduce latency and cloud load. \\
%		\hline
%	\end{tabular}
%	\caption{Representative Pico-Cloud use case categories with brief descriptions.}
%	\label{table:use_cases}
%\end{table*}

\subsection{Edge AI Inference}
Edge AI inference with Pico-Cloud enables the execution of streamlined artificial intelligence algorithms directly on compact edge devices~\cite{singh2023edge}. This capability is particularly valuable in environments that require immediate responses or lack stable connectivity~\cite{shi2020communication}, as it minimizes latency and reduces reliance on external computing resources.

\subsection{Rural Connectivity}
In remote and rural areas, Pico-Cloud supports the delivery of essential local digital services such as community messaging platforms, educational repositories, and public data access points. It improves digital inclusivity by overcoming infrastructural limitations associated with conventional internet connectivity~\cite{helmer2016bringing}.

\subsection{Educational Labs}
Educational institutions adopt Pico-Cloud to provide hands-on training in cloud computing, distributed systems, containerization, and DevOps. These affordable and scalable lab environments significantly reduce costs and technical barriers, promoting equitable access to contemporary computing skills~\cite{xu2013cloud}.

\subsection{Disaster Recovery}
In post-disaster contexts, Pico-Cloud solutions offer rapidly deployable, resilient, and energy-efficient infrastructure for restoring essential communication networks and local IT services. Their portability and low power requirements make them particularly suitable for emergency response and humanitarian aid operations~\cite{abualkishik2020disaster}.

\subsection{Local Private Cloud}
For privacy-conscious individuals and organizations, Pico-Cloud provides a secure, self-managed platform for hosting personal data storage, web applications, and IoT management systems~\cite{davenport2019air}. This decentralization strengthens data sovereignty, reduces dependence on commercial cloud providers, and mitigates privacy risks.

\subsection{Innovation in Underserved Areas}
Pico-Cloud enables digital services in resource-limited environments such as off-grid villages, field labs, and remote research stations~\cite{BorneoAudio2020,Correa2024Leaf}. Its low power footprint and portability support deployments in locations without reliable electricity or connectivity, facilitating local data processing, educational access, and grassroots innovation.

\subsection{IoT Edge Gateway}
When used as an IoT edge gateway, Pico-Cloud enables local aggregation, filtering, and analysis of sensor-generated data. By performing preliminary processing at the edge, these gateways reduce bandwidth usage, lower response times, and improve the overall efficiency of IoT deployments~\cite{condry2016using}.

\section{Related Work}
Recent developments in edge computing and lightweight cloud infrastructure have demonstrated the growing importance of decentralized, low-power platforms similar in spirit to the Pico-Cloud. Elkhatib et al.~\cite{elkhatib2017microclouds} explored the feasibility of micro-clouds built from Raspberry Pi devices to host fog applications under constrained network conditions. Their experimental evaluations showed that such platforms can support web services and application delivery at the edge.

In 2021, the \textit{Con-Pi} framework~\cite{rahman2021conpi} introduced a container-based edge orchestration system using clustered Raspberry Pi units. The platform was validated in agricultural automation tasks, including pest deterrence and remote sensing, where latency and resilience were critical. These systems illustrate the utility of Pico-Cloud concepts in real-time and field-deployed AI applications.

Industrial applications have likewise leveraged single-board computers (SBCs) as edge gateways. A study on SBC-powered smart grid monitoring~\cite{gheisari2022raspberry} demonstrated real-time on-site telemetry processing, reducing reliance on upstream cloud infrastructure. This aligns with Pico-Cloud’s emphasis on local computation and bandwidth efficiency.

In the domain of edge AI, quantized deep learning models (e.g., YOLOv4-Tiny) have been successfully deployed on Raspberry Pi 4 and Pi 5 devices~\cite{hossain2023real}, showing that modern AI workloads can operate within the constrained computational envelope of Pico-Cloud hardware.

Finally, NodeWeaver~\cite{NodeWeaver2023} and Canonical’s MicroCloud~\cite{canonical2023microcloud} exemplify industrial-grade platforms that bring distributed cloud services to the network edge. While these efforts often target higher-end SBCs or compact servers, they reinforce the architectural motivation behind Pico-Clouds: lightweight, self-contained, and locally managed cloud stacks.

Taken together, prior work highlights a growing trend toward lightweight, decentralized infrastructures, but most solutions either target specialized applications or depend on comparatively powerful devices. In contrast, the Pico-Cloud focuses on extremely resource-constrained, widely accessible platforms, aiming to provide a systematic architecture for affordable, portable, and sustainable cloud-native services.

\section{Design and Implementation}
Pico-Cloud deployments require a tightly scoped technical architecture to operate effectively within the severe constraints of ultra-compact hardware. Unlike conventional edge or cloud systems, Pico-Cloud nodes typically lack advanced networking, memory, or storage subsystems. As such, design choices must prioritize minimal resource consumption, energy efficiency, and component interoperability. This section outlines the key elements of a functional Pico-Cloud platform, including hardware selection, networking configuration, software stack, orchestration strategies, and power management.

\begin{figure}[t]
	\centering
	\includegraphics[width=\linewidth]{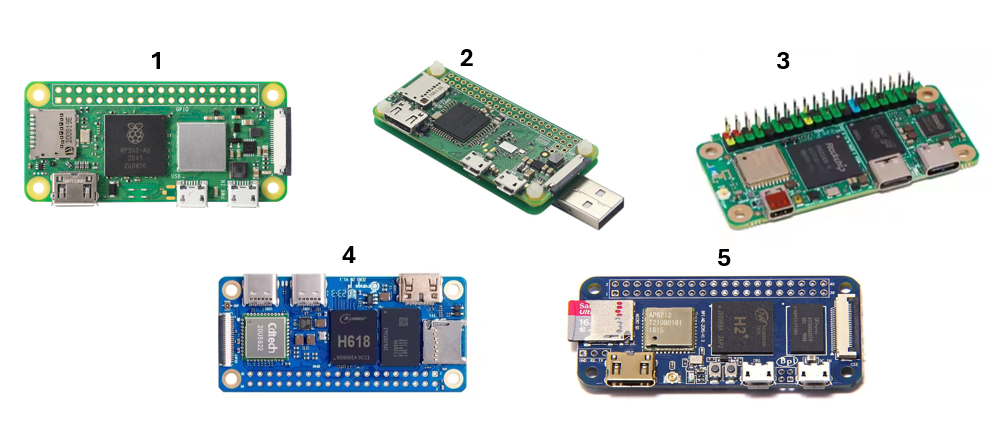}
	\caption{Representative ultra-compact single-board computers (SBCs) for Pico-Cloud deployments: 
		Raspberry Pi Zero 2 W (top-left), Raspberry Pi Zero (top-center), Radxa Zero (top-right), Orange Pi Zero 2 (bottom-left), and Banana Pi M2 Zero (bottom-right).}
	\label{fig:sbc-gallery-five}
\end{figure}

\subsection{Hardware Considerations}
The hardware foundation of a Pico-Cloud is defined by its reliance on low-cost, low-power, small-footprint devices capable of running essential cloud-like workloads. The design objective is to support distributed computation, lightweight virtualization, and reliable service delivery using micro-scale resources, particularly in locations where traditional infrastructure is infeasible or uneconomical.

At the core of most implementations are single-board computers (SBCs) such as the Raspberry Pi Zero~\cite{RaspberryPiZero} and Zero 2 W~\cite{RaspberryPiZero2W}. These devices provide adequate computational capability—ranging from a 1\,GHz ARM11 processor with 512\,MB RAM to a quad-core ARM Cortex-A53—while operating within a thermal design power (TDP) below 3\,W. Their compact form factor, low idle power draw, and support for mainstream Linux distributions make them well-suited for embedded deployments. The Raspberry Pi Zero’s USB gadget mode further enables networking and power delivery over USB, reducing the need for external switches or power circuits. This feature is frequently leveraged in cluster backplane solutions such as the ClusterHAT~\cite{ClusterHAT}.

Alternative SBC platforms, including the Orange Pi Zero~\cite{OrangePiZero}, Banana Pi M2 Zero~\cite{BananaPiM2Zero}, and Radxa Zero~\cite{RadxaZero}, provide similar performance envelopes with variations in I/O and wireless features. Platform choice can be tailored to application requirements such as GPIO availability, supported operating systems, or cost-per-unit optimization.

For scenarios requiring greater computational capacity—such as workload coordination, model inference, or persistent data services—higher-tier SBCs like the Raspberry Pi 4B (1–8\,GB RAM)~\cite{RaspberryPi4B}, NVIDIA Jetson Nano~\cite{JetsonNano}, or ROCK Pi series~\cite{ROCKPi} can be integrated into heterogeneous Pico-Cloud clusters. These more capable nodes act as orchestrators, caching endpoints, or AI accelerators while maintaining overall system efficiency and modularity.

\begin{figure}[t]
	\centering
	\includegraphics[width=0.5\linewidth]{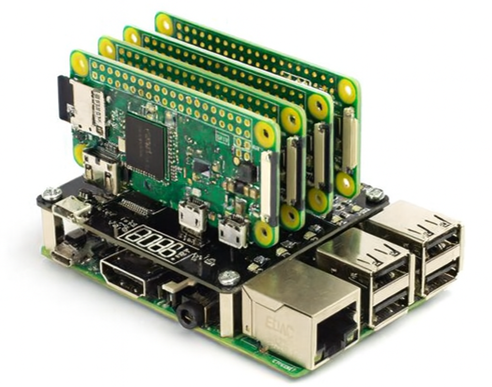}
	\caption{ClusterHAT hosting four Raspberry Pi Zero boards with consolidated USB power and networking for compact Pico-Cloud deployments.}
	\label{fig:clusterhat}
\end{figure}

\subsection{Thermal and Energy}
Thermal design is typically passive due to the low wattage of individual nodes. However, in denser clusters or warmer environments, passive heat sinks or low-profile fans may be required to prevent throttling. Physical layouts range from loosely coupled nodes to stackable cluster boards such as the ClusterHAT~\cite{ClusterHAT}, which consolidates power and USB networking for up to four Raspberry Pi Zero devices through a single USB hub and GPIO interface.

Energy independence is a key enabler of Pico-Cloud practicality. Most nodes operate within a 0.7–2.5\,W envelope, allowing battery-powered or solar-backed deployments in off-grid environments. USB power banks, lithium-ion modules with charge controllers, or solar microgrids can sustain long-term operation without AC power. This power profile supports deployments in humanitarian relief, rural connectivity, and mobile edge infrastructures.

The total cluster power consumption can be approximated as

\begin{equation}
	P_{\mathrm{cluster}} = \sum_{i=1}^{N} P_i + P_{\mathrm{hub}} + P_{\mathrm{overhead}},
	\label{eq:cluster-power}
\end{equation}

where $P_i$ is the average draw of node $i$, $P_{\mathrm{hub}}$ the consumption of the backplane or hub, and $P_{\mathrm{overhead}}$ any additional regulators or cooling elements. The expected runtime on a given battery is estimated by

\begin{equation}
	T_{\mathrm{runtime}} \approx \frac{E_{\mathrm{battery}} \cdot \eta}{P_{\mathrm{cluster}}},
	\label{eq:battery-runtime}
\end{equation}

where $E_{\mathrm{battery}}$ is the battery capacity in watt-hours and $\eta$ the conversion efficiency. These approximations provide practical guidance for sizing mobile or solar-backed deployments.

\subsection{Network Configuration}
Networking within Pico-Cloud infrastructures presents unique challenges due to the limited I/O capabilities of ultra-compact SBCs, many of which lack native Ethernet interfaces. Effective network design must therefore minimize hardware overhead while ensuring interoperability and resilience.

In Raspberry Pi Zero–based clusters, the most common solution is USB gadget mode, which creates Ethernet-over-USB interfaces. Each Zero connects to a host controller—typically a Raspberry Pi 3 or 4—via USB OTG. The controller acts as a hub and bridge, assigning virtual interfaces and facilitating inter-node communication. This approach not only supports data transmission but also powers the connected nodes through a unified interface, reducing cabling complexity.

Cluster backplane solutions such as the ClusterHAT and PiTray Mini~\cite{PiTrayMini} further simplify topologies by integrating USB routing and GPIO-based power control. These designs allow up to four Raspberry Pi Zero devices to be powered, monitored, and networked without external switches. USB~2.0 gadget networks typically sustain transfer rates of 8–10\,MB/s, sufficient for low-throughput workloads.

Where USB networking is impractical, onboard Wi-Fi (e.g., Raspberry Pi Zero W, Orange Pi Zero 2) enables ad-hoc mesh or AP-based topologies. While simplifying deployment, wireless links introduce challenges with interference, reliability, and multicast service discovery (e.g., mDNS/Avahi~\cite{RFC6762,RFC6763}). Careful channel planning and isolation are therefore necessary.

For higher throughput, wired Ethernet remains preferable when supported. Devices such as the Raspberry Pi 4 or ROCK Pi series provide Gigabit Ethernet and can serve as aggregation nodes or uplinks in star or tree topologies. These configurations are well-suited to use cases involving persistent data streams such as IoT telemetry, video processing, or federated learning.

Across all topologies, service discovery and address assignment rely on lightweight mechanisms. Static IPs are common in isolated clusters, while DHCP or DNS-based registration is used in dynamic deployments. In Kubernetes-managed clusters (e.g., with k3s), overlay solutions such as Flannel~\cite{Flannel} or WireGuard~\cite{WireGuard} enable secure service routing across nodes.

The effective throughput of Pico-Cloud interconnects can be approximated as

\begin{equation}
	C_{\mathrm{eff}} \approx C_{\mathrm{phy}} \cdot (1 - \alpha_{\mathrm{proto}} - \alpha_{\mathrm{contention}}),
	\label{eq:net-capacity}
\end{equation}

where $C_{\mathrm{phy}}$ is the physical link rate (e.g., 480\,Mb/s for USB~2.0, 150--300\,Mb/s for 802.11n, or 1\,Gb/s for Ethernet), $\alpha_{\mathrm{proto}}$ the protocol overhead, and $\alpha_{\mathrm{contention}}$ the factor due to medium contention and interference. In practice, this corresponds to sustained rates of 8–10\,MB/s for USB gadget links, tens of MB/s for Wi-Fi, and near line-rate for wired Gigabit Ethernet.

\subsection{Software Stack}
The software architecture of the Pico-Cloud platform is designed to provide cloud-like functionality while operating under the stringent resource limitations of ultra-compact hardware. Rather than replicating the full feature set of large-scale platforms, Pico-Cloud software stacks focus on essential primitives—such as isolation, deployment, orchestration, and networking—within a minimal and lightweight system.

\textbf{Operating System Layer.}
At the foundation, Pico-Cloud nodes run headless Linux distributions optimized for small memory and storage footprints. Common choices include Raspberry Pi OS Lite~\cite{RaspberryPiOS}, Ubuntu Server for ARM~\cite{UbuntuServerARM}, Alpine Linux~\cite{AlpineLinux}, and DietPi~\cite{DietPi}. These distributions typically boot with under 100\,MB of RAM and occupy less than 1\,GB of disk space. Kernel configurations are tailored to support containerization (via cgroups, namespaces, overlayfs~\cite{LinuxNamespaces,OverlayFS}) while omitting unnecessary drivers or UI components.

\textbf{Container Runtime and Isolation.}
Application isolation is achieved through Linux containers~, which offer a lightweight alternative to full virtualization. Docker~ remains the most widely adopted runtime due to its ARM support, mature ecosystem, and active development. Alternatives such as Podman~\cite{Podman} or containerd are used where lower overhead or rootless execution is required. Containers are typically based on minimal images (e.g., Alpine~\cite{AlpineLinux}, BusyBox~\cite{BusyBox}) to conserve resources. ARM32/ARM64 compatibility is maintained through architecture-specific builds and multi-arch container manifests.

\textbf{Service Orchestration.}
Lightweight orchestrators coordinate workloads across Pico-Cloud nodes. K3s~\cite{K3s}, a compact Kubernetes distribution, reduces the Kubernetes control plane to a single binary and supports clustering on devices with as little as 512\,MB RAM. When Kubernetes overhead is prohibitive, HashiCorp Nomad~\cite{Nomad} offers a single-agent model with optional integration with Consul~\cite{Consul} for service registration and health checks. Docker Swarm~\cite{DockerSwarm} provides a simpler alternative for container clustering without additional dependencies.

\textbf{Service Discovery and Networking.}
Internal service resolution and communication rely on a mix of lightweight mechanisms. In small USB- or Wi-Fi–based clusters, local mDNS discovery via Avahi~\cite{Avahi} or systemd-resolved~\cite{SystemdResolved} is sufficient. More structured environments employ Consul~\cite{Consul} or Kubernetes-native DNS~\cite{K8sDNS}. Overlay networks enabled by lightweight CNI plugins such as Flannel~\cite{Flannel} or WireGuard tunnels~\cite{WireGuard} provide secure, isolated communication across nodes.

\textbf{Monitoring and Observability.}
Monitoring in Pico-Clouds must balance visibility with overhead. Node metrics can be collected using Prometheus Node Exporter~\cite{Prometheus} or Telegraf~\cite{Telegraf}, while logging is often handled by journald~\cite{SystemdJournal}, logrotate~\cite{Logrotate}, or lightweight log collectors such as Fluent Bit~\cite{FluentBit}. When dashboards are needed, visualization tools like Grafana~\cite{Grafana} are hosted on a more capable node or external server.

\textbf{Security and Access Control.}
Security is addressed with minimal but essential mechanisms. SSH with key-based authentication~\cite{OpenSSH} is standard, supported by firewall rules using iptables or nftables~\cite{Iptables,Nftables}. Containers can be sandboxed with Seccomp~\cite{Seccomp} and AppArmor~\cite{AppArmor}. For sensitive deployments, VPN tunnels (e.g., WireGuard~\cite{WireGuard}), secret storage (Vault~\cite{Vault}), and non-root container execution are selectively introduced according to the threat model.

\subsection{Storage and Data Architecture}
Storage in Pico-Cloud deployments must balance capacity, persistence, and performance within the physical and thermal constraints of low-power, flash-based devices. Whereas traditional cloud systems rely on high-throughput block storage or distributed file systems, Pico-Cloud nodes typically use microSD cards or low-profile flash drives with limited endurance and bandwidth. This necessitates storage approaches that are minimal yet resilient to failure and wear.

\textbf{Local Storage Per Node.}
Each node typically maintains a local filesystem on a microSD card, which holds the operating system, container engine, and configuration data. Because flash media is subject to limited write cycles, common practices include mounting ephemeral or high-churn directories (e.g., `/var/log`, container overlay layers) as RAM disks (`tmpfs`) or relocating them to more durable external storage. File system integrity tools (e.g., fsck, overlayfs~\cite{OverlayFS}) are used to mitigate corruption following power loss or improper shutdowns.

\textbf{Shared or Centralized Storage.}
In multi-node deployments, a more capable node—often with USB-attached SSD or eMMC—can serve as a storage provider via lightweight protocols such as NFS~\cite{NFS} or Samba~\cite{Samba}. This enables distributed nodes to access persistent volumes for logs, databases, or shared applications without duplicating data locally. However, such configurations introduce a potential single point of failure unless redundancy or snapshotting mechanisms are applied.

\textbf{Distributed File Systems.}
Advanced clusters may employ distributed storage systems such as GlusterFS~\cite{GlusterFS}, MooseFS~\cite{MooseFS}, or IPFS~\cite{IPFS}. While these platforms offer redundancy and scalability, their resource demands often exceed the capabilities of entry-level Pico-Cloud nodes. In practice, they are usually confined to deployments involving higher-end SBCs (e.g., Raspberry Pi 4B or Jetson Nano) or hybrid clusters with centralized coordination. Lightweight distributed key-value stores such as etcd~\cite{etcd} or Redis~\cite{Redis} (in append-only mode) are sometimes used for configuration replication or buffering across the cluster.

\textbf{Data Volatility and Ephemeral Design.}
Given power variability and limited durability of flash media, Pico-Cloud applications are frequently designed to operate in an ephemeral or stateless mode. Services avoid continuous writes and, where persistence is required, synchronize critical data to off-cluster storage or cloud backends when connectivity permits. This design philosophy aligns with an “eventual durability” model, where transient state may be lost but essential datasets are periodically consolidated upstream.

\textbf{Data Synchronization and Backup.}
When external connectivity is available, lightweight synchronization tools such as \texttt{rsync}~\cite{rsync}, \texttt{rclone}~\cite{rclone}, or \texttt{syncthing}~\cite{Syncthing} are often used to replicate logs, configuration files, or application output. Synchronization is typically scheduled for off-peak periods or when power and bandwidth conditions are favorable. Incremental snapshotting and differential sync strategies help reduce I/O wear and network congestion.

Overall, Pico-Cloud storage architectures are shaped less by full-scale cloud paradigms than by constraints of endurance, power, and coordination overhead. Selective persistence, lightweight replication, and fail-soft design patterns allow the system to remain functional and resilient while staying within the limits of small-form-factor hardware.

\subsection{Security Considerations}
Given the resource constraints and deployment contexts of Pico-Cloud systems, security must be approached with a lightweight yet robust mindset. These systems are often deployed in unmonitored or adversarial environments, operate autonomously, and lack the extensive control surfaces typical of enterprise infrastructure. Their security model therefore emphasizes minimal attack surface, strong defaults, and simple, auditable mechanisms for access control and data integrity.

\textbf{Authentication and Access Control.}
Pico-Cloud nodes are typically managed via secure shell (SSH)~\cite{OpenSSH} with public key authentication. Password logins are disabled by default to reduce brute-force exposure. Role separation is enforced by creating non-root administrative accounts and limiting `sudo` privileges. In multi-node clusters, automated mechanisms such as rotating SSH keys or ephemeral credentials are recommended for scalable orchestration.

\textbf{Network Security.}
Network interfaces are tightly firewalled using iptables~\cite{Iptables} or nftables~\cite{Nftables}, with only required ports exposed. Internal communication can be secured using lightweight VPN tunnels (e.g., WireGuard~\cite{WireGuard}) or mutual TLS. For external connectivity with upstream cloud or APIs, outbound requests are filtered and logged to reduce leakage risks.

\textbf{Container Isolation.}
Containers are executed with least privilege using rootless runtimes such as Docker (rootless mode)~\cite{Docker} or Podman~\cite{Podman}. Linux namespaces~\cite{LinuxNamespaces}, cgroups, and Seccomp~\cite{Seccomp} profiles limit system call access and filesystem exposure. SELinux~ or AppArmor~\cite{AppArmor}, where supported, provide additional workload sandboxing.

\textbf{Update and Patch Management.}
Limited or intermittent connectivity makes continuous patch pipelines impractical. Instead, updates are applied via offline-signed bundles or controlled waves using cluster-aware tools such as Ansible~\cite{Ansible}, k3sup~\cite{k3sup}, or Syncthing~\cite{Syncthing}. Integrity is verified with checksums or PGP signatures, and rollback snapshots are configured for recovery from failed updates.

\textbf{Secrets Management.}
Secrets such as API tokens or encryption keys are stored in encrypted volumes or injected at runtime via environment variables. Tools such as \texttt{sops}~\cite{sops}, \texttt{age}~\cite{age}, or Vault Agent~\cite{Vault} support distributed secret rotation in clustered deployments. On single-node setups, secrets may be embedded in immutable containers or stored on removable media with physical access control.

\textbf{Resilience to Physical Compromise.}
Because many Pico-Cloud nodes operate unattended, physical protection is critical. Devices may be configured with encrypted root filesystems (e.g., LUKS~\cite{LUKS}), while secure boot features (e.g., Raspberry Pi 4 EEPROM~\cite{Pi4Boot}) enforce integrity at startup. On more constrained devices, read-only root partitions and simple tamper indicators (e.g., GPIO-based alerts) offer baseline protection.

The Pico-Cloud security model adopts a layered defense strategy aligned with its minimalist philosophy. Instead of replicating enterprise-grade frameworks, it relies on a curated set of simple, transparent, and verifiable controls that preserve resilience and trustworthiness even in challenging deployment environments.

\begin{figure}[t]
	\centering
	\includegraphics[width=0.6\linewidth]{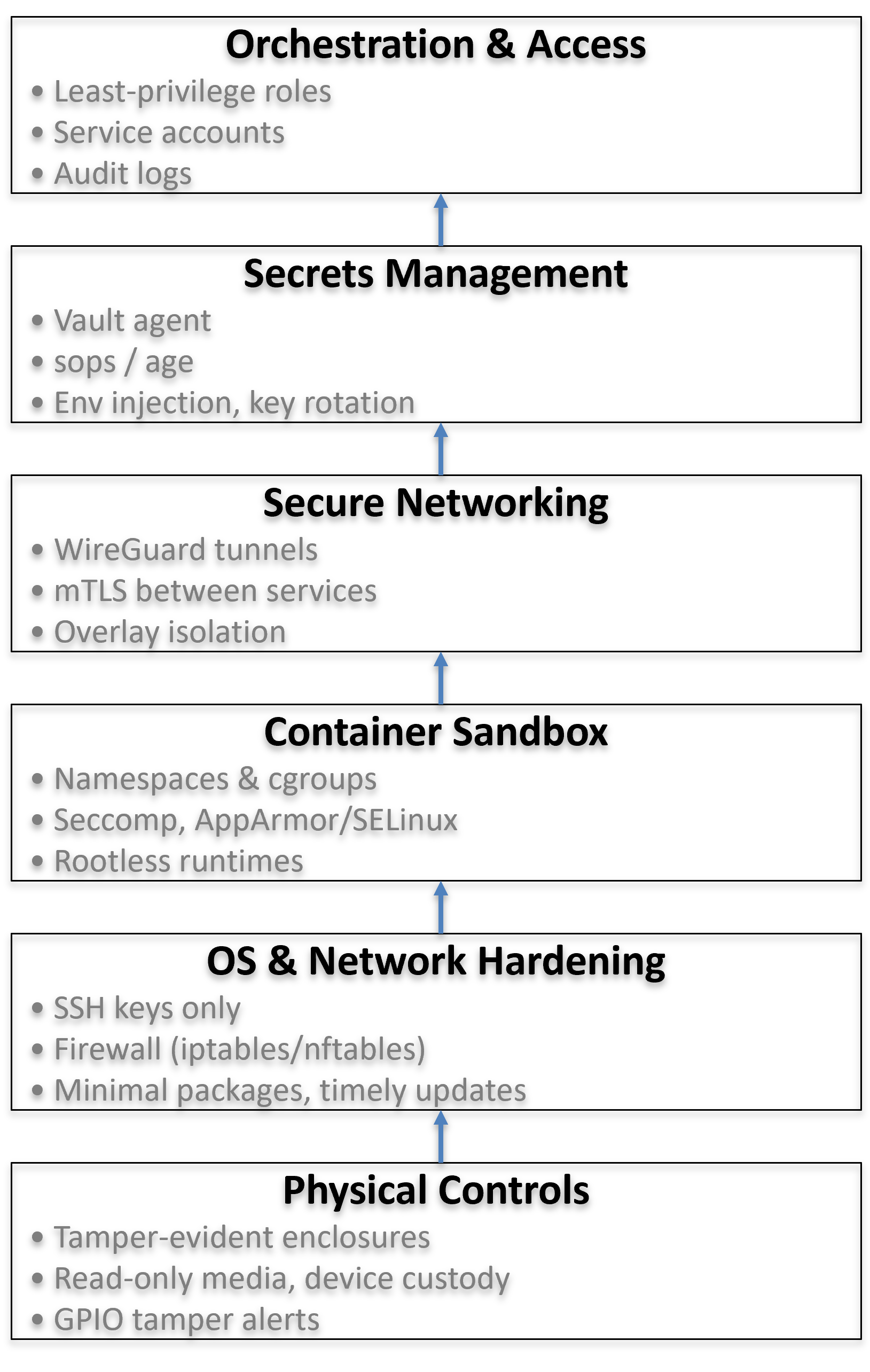}
	\caption{Defense-in-depth for Pico-Clouds, layered from physical controls to orchestration and access. Arrows indicate increasing assurance with stacked protections.}
	\label{fig:defense-in-depth}
\end{figure}

“Figure~\ref{fig:defense-in-depth} summarizes the Pico-Cloud defense-in-depth model, from physical controls through OS/network hardening, container sandboxing, secure networking, secrets management, and orchestration/access.”

\section{Deployment Models}
Pico-Cloud infrastructures support a range of deployment models, each tailored to different operational scenarios, constraints, and levels of system complexity. Unlike conventional data center architectures, these deployments are explicitly optimized for minimalism, resilience, and autonomy. This subsection outlines the principal deployment topologies used in real-world Pico-Cloud configurations.

\textbf{Single-Node Cloudlet.}  
The simplest form of deployment is a single-node cloudlet, wherein all core services are hosted on a standalone single-board computer. This model is well suited to portable or embedded applications such as mobile sensing, localized media servers, or home automation. Despite its lack of redundancy, it provides self-contained functionality with minimal hardware and energy overhead, making it ideal for isolated environments with intermittent or no network access.

\textbf{Static Micro-Cluster.}  
A small group of homogeneous SBCs can be co-located to form a static micro-cluster. Nodes are networked via USB or Wi-Fi and may perform distributed roles, such as separation of application services, persistent storage, or telemetry collection. Container orchestration frameworks such as K3s or Docker Swarm facilitate service scheduling and resource balancing. This model increases availability and functional distribution without significantly raising complexity or power demands.

\textbf{Heterogeneous Edge Cluster.}  
In mixed-resource deployments, heterogeneous clusters combine low-power nodes (e.g., Pi Zero) with higher-capability orchestrators (e.g., Raspberry Pi 4B or NVIDIA Jetson Nano). This topology allows for differentiated service roles—such as inference, caching, and gateway functionality—while preserving the lightweight and modular ethos of Pico-Clouds. It is applicable to smart buildings, industrial monitoring, and AI-enabled edge workloads where role specialization improves performance.

\textbf{Mobile and Field Deployments.}  
Pico-Cloud clusters may be embedded into mobile enclosures, field kits, or transportable nodes for use in remote sensing, disaster recovery, or environmental monitoring. These deployments emphasize portability, ruggedness, and autonomy. Services are often configured for local caching or asynchronous synchronization, enabling continuous functionality despite intermittent connectivity. Power is typically supplied by batteries or solar microgrids, reinforcing suitability for off-grid contexts.

\textbf{Federated and Mesh Architectures.}  
In distributed and decentralized environments, multiple Pico-Cloud units may participate in federated or peer-to-peer architectures. Each node or sub-cluster operates semi-autonomously while coordinating workloads or data streams over mesh Wi-Fi or ad-hoc networking. These topologies are valuable in civic networks, educational mesh labs, or rural infrastructures. Lightweight service discovery and eventual consistency mechanisms synchronize state without centralized control.

\textbf{Hybrid Cloud Extensions.}  
Pico-Cloud systems may also integrate into hybrid models wherein local edge services operate in tandem with public cloud infrastructure. Nodes act as pre-processing endpoints or fallback service replicas, reducing dependence on wide-area connectivity. Applications include bandwidth optimization, offline-first services, and gateway nodes in larger IoT systems. While this approach extends reliability through cloud integration, it adds WAN dependency and integration overhead.

Each of these models presents trade-offs across dimensions such as fault tolerance, orchestration complexity, latency, and deployment cost. Designers are encouraged to align architectural choices with workload demands and environmental constraints to fully leverage the strengths of the Pico-Cloud paradigm. A comparative overview of these deployment models is summarized in Table~\ref{tab:deployment-models}.

\begin{table*}[t]
	\centering
	\caption{Summary of Pico-Cloud Deployment Models}
	\begin{tabular}{|p{3.2cm}||p{4.5cm}|p{4.5cm}|p{4.5cm}|}
		\hline
		\textbf{Model} & \textbf{Description} & \textbf{Use Cases} & \textbf{Key Trade-offs} \\
		\hline
		Single-Node Cloudlet & Standalone SBC hosting all services. Minimal hardware and energy overhead. & Mobile sensing, home automation, local media servers, isolated/offline environments. & Very low complexity and cost, but no redundancy or fault tolerance. \\
		\hline
		Static Micro-Cluster & Homogeneous SBCs co-located, connected via USB/Wi-Fi; containers orchestrated with K3s or Docker Swarm. & Small-scale service separation, persistent storage, telemetry collection. & Improved availability and resource separation, modest power/cost increase. \\
		\hline
		Heterogeneous Edge Cluster & Mix of low-power nodes (e.g., Pi Zero) with higher-tier orchestrators (e.g., Pi 4B, Jetson Nano). Specialized service roles (inference, caching, gateways). & Smart buildings, industrial monitoring, AI-enabled edge workloads. & Balances efficiency with performance, requires orchestration of mixed hardware. \\
		\hline
		Mobile and Field Deployments & Clusters embedded in portable kits/enclosures; powered by batteries or solar. Services often designed for offline operation with sync. & Disaster recovery, environmental monitoring, remote sensing, off-grid scenarios. & Rugged and autonomous but limited by intermittent connectivity and energy availability. \\
		\hline
		Federated and Mesh Architectures & Multiple clusters/nodes coordinated in peer-to-peer or mesh Wi-Fi networks. Eventual consistency without central coordination. & Civic networks, rural connectivity, educational mesh labs. & Decentralized and resilient, but weaker global coordination and consistency. \\
		\hline
		Hybrid Cloud Extensions & Local Pico-Clouds integrated with public cloud; act as preprocessing, replicas, or fallback. & IoT gateways, bandwidth optimization, offline-first apps. & Extends reliability with cloud support but adds WAN dependency and integration overhead. \\
		\hline
	\end{tabular}
	\label{tab:deployment-models}
\end{table*}

\section{Performance Considerations}
Assessing the practicality of Pico-Cloud deployments requires considering known performance characteristics of single-board computers and lightweight cloud stacks, rather than conducting exhaustive benchmarking. Since these systems are highly resource-constrained, the relevant question is not peak throughput but whether they can provide service responsiveness and energy efficiency sufficient for edge-centric workloads. This section outlines expected performance profiles based on vendor specifications and evaluations reported in prior work.

\textbf{Boot and Service Initialization.}  
Cold-boot times for devices such as the Raspberry Pi Zero 2~W are typically on the order of tens of seconds when using lightweight Linux distributions (e.g., Raspberry Pi OS Lite). This enables nodes to recover quickly from power interruptions, which is valuable in duty-cycled or energy-aware deployments.

\textbf{Container Startup Time.}  
Evaluations of container runtimes on ARM-based SBCs consistently show that minimal images (e.g., Alpine or BusyBox) can be instantiated with very low overhead, often in less than a second. Startup latency increases when services require heavier initialization (e.g., Python or Node.js applications), but remains acceptable for edge workloads where rapid elasticity is not the primary goal.

\textbf{CPU and Memory Utilization.}  
Prior studies and community benchmarks highlight memory as the most significant bottleneck for devices with 512\,MB RAM or less. The operating system kernel and background services consume a baseline footprint, leaving limited headroom for containerized applications. CPU performance, however, is generally sufficient for lightweight workloads such as serving static content, telemetry collection, or small-scale inference with frameworks like TensorFlow Lite.

\textbf{Power Draw and Thermal Envelope.}  
According to manufacturer specifications and independent testing, SBCs such as the Raspberry Pi Zero and Zero 2~W typically operate in a 1–3\,W power envelope. This allows clusters of several nodes to remain under 10\,W in aggregate. Passive cooling is generally adequate, as these boards rarely exceed thermal limits under sustained load in ventilated environments.

\textbf{Network Throughput and Latency.}  
Networking performance depends strongly on the chosen interconnect. USB gadget networking provides modest throughput (typically 8–10\,MB/s), which is sufficient for control-plane traffic and small-scale data sharing. Wireless deployments are more variable, with throughput and latency influenced by interference and signal strength. Wired Ethernet on higher-tier boards (e.g., Raspberry Pi~4B) offers more reliable gigabit-class connectivity.

\textbf{Availability and Fault Tolerance.}  
Lightweight orchestrators such as k3s and Docker Swarm can tolerate node churn and service restarts, although features like live migration or consensus protocols are limited by hardware constraints. Fault tolerance in Pico-Cloud clusters is therefore achieved through container redeployment and lightweight replication strategies rather than full high-availability semantics.

Taken together, these observations show that Pico-Cloud systems are not designed for enterprise-class throughput or latency guarantees. Their value lies in demonstrating that even severely resource-constrained nodes can deliver cloud-like behaviors—such as container orchestration, service elasticity, and multi-node coordination—at very low cost and energy consumption. Performance should therefore be viewed as “good enough” for disconnected operations, local data aggregation, and educational or prototyping contexts, rather than as a replacement for conventional edge or cloud servers. A consolidated overview of these characteristics is provided in Table~\ref{tab:performance}, which summarizes the trade-offs across dimensions such as boot latency, container overhead, resource utilization, and fault tolerance.

\begin{table*}[t]
	\centering
	\caption{Summary of Expected Performance Characteristics of Pico-Cloud Nodes}
	\begin{tabular}{|p{3.2cm}||p{5cm}|p{7cm}|}
		\hline
		\textbf{Dimension} & \textbf{Typical Characteristics} & \textbf{Notes and Implications} \\
		\hline
		Boot and Initialization & Tens of seconds (e.g., Pi Zero 2~W with Raspberry Pi OS Lite) & Suitable for duty-cycled or energy-aware deployments where rapid restart is valuable. \\
		\hline
		Container Startup & Sub-second for minimal images (Alpine, BusyBox); longer for heavier runtimes (Python, Node.js) & Low overhead for lightweight services; still acceptable for edge workloads with modest elasticity needs. \\
		\hline
		CPU and Memory Utilization & CPU sufficient for web serving, telemetry, and small ML inference; RAM is critical bottleneck (512\,MB typical) & Careful orchestration needed; avoid large in-memory caches or memory-fragmenting tasks. \\
		\hline
		Power and Thermal Envelope & 1–3\,W per node; clusters of 4 nodes under 10\,W; passive cooling generally sufficient & Enables battery- or solar-backed operation; thermal throttling rare under ventilated conditions. \\
		\hline
		Network Throughput and Latency & USB gadget networking: $\sim$8–10 MB/s; Wi-Fi variable; higher-tier boards provide Gigabit Ethernet & Adequate for control-plane traffic, telemetry, and modest data sharing; wired preferred for reliability. \\
		\hline
		Availability and Fault Tolerance & Orchestrators (k3s, Docker Swarm) handle node churn and redeployment; no live migration & Fault tolerance achieved via container redeployment and lightweight replication rather than full HA semantics. \\
		\hline
	\end{tabular}
	\label{tab:performance}
\end{table*}

\section{Challenges and Limitations}
Despite the versatility and accessibility of Pico-Clouds, their minimalist design introduces a series of technical and operational constraints that must be considered in any deployment. These limitations reflect trade-offs made to support ultra-low-cost, ultra-low-power computing at the network edge.

\textbf{Limited Processing and Memory Capacity.}  
Pico-Cloud nodes, such as Raspberry Pi Zero-class devices, are equipped with modest single-core or low-power ARM processors and typically offer only 512\,MB to 1\,GB of RAM. These limitations severely restrict workloads that require high concurrency, in-memory databases, or large-scale computation. Even with parallelization, performance gains remain marginal due to communication overhead and limited per-node speed~\cite{elkhatib2017microclouds}.

\textbf{Storage and I/O Bottlenecks.}  
Most Pico-Cloud nodes rely on microSD cards or USB flash drives, which are slow, wear-prone, and capacity-limited. This makes them unsuitable for write-heavy workloads or persistent database operations. Storage redundancy, such as RAID or distributed file systems, is difficult to implement efficiently on these platforms.

\textbf{Bandwidth and Network Latency.}  
Due to reliance on USB gadget interfaces or shared Wi-Fi networks, Pico-Cloud clusters often suffer from low throughput (on the order of 100\,Mbps) and increased latency. This limits their suitability for real-time communication, high-volume data transfers, or multi-client streaming applications. Wireless deployments are especially prone to interference, packet loss, and degraded cluster coordination.

\textbf{Scalability and Coordination Overhead.}  
While Pico-Clouds are horizontally scalable in principle, scaling beyond a few nodes quickly introduces coordination and management complexity. Scheduling, health monitoring, and service discovery must remain lightweight, or the overhead will overwhelm node resources. Furthermore, scaling does not lead to linear performance improvements due to I/O and memory bottlenecks~\cite{rahman2021conpi}.

\textbf{Security and Physical Exposure.}  
Deployed in open or unmonitored environments, Pico-Cloud nodes are vulnerable to physical tampering, theft, and data exfiltration. Software security is also constrained: few boards support secure boot or hardware-backed key storage. Without proper SSH key management and network isolation, clusters remain at risk from unauthorized access.

\textbf{Maintainability and Fault Recovery.}  
Each node is an independent unit with its own OS and container runtime, which introduces overhead in updates, logging, and recovery. Unlike server-grade systems, Pico-Clouds lack remote power cycling, serial consoles, or out-of-band management, complicating failure response. SD card corruption due to power loss remains a common failure mode~\cite{NodeWeaver2023}.

\textbf{Software Compatibility.}  
Because many off-the-shelf enterprise applications are designed for x86 or high-memory architectures, users may need to recompile software, optimize configurations, or choose lighter alternatives (e.g., SQLite instead of PostgreSQL, Nginx instead of Apache). This limits compatibility with commercial cloud-native platforms and managed services.

\begin{table*}[t]
	\centering
	\caption{Summary Comparison Across Deployment Models}
	\label{table:comparison}
	\begin{tabular}{|l||c|c|c|}
		\hline
		\textbf{Attribute} & \textbf{Cloud} & \textbf{Edge} & \textbf{Pico-Cloud} \\
		\hline
		Scale & Massive & Medium & Micro \\
		\hline
		Hardware & Data center-grade & Industrial/SBC & Low-power SBCs \\
		\hline
		Power Use & High & Moderate & Very Low \\
		\hline
		Connectivity & Always-on Internet & WAN/LAN Required & Offline-capable \\
		\hline
		Deployment & Centralized & Regionally distributed & Localized, manual \\
		\hline
		Security Model & Multi-tenant, IAM & Role-based, some IAM & Single-tenant, containerized \\
		\hline
		Automation & Full orchestration & Partial orchestration & Scripted or manual \\
		\hline
		Main Use Cases & Web-scale, ML, SaaS & IoT, AI inference, CDN & Education, field ops, autonomy \\
		\hline
		Cost Profile & High CapEx/OpEx & Medium CapEx & Very Low CapEx \\
		\hline
	\end{tabular}
\end{table*}

		\section{Comparison with Traditional Cloud and Edge Computing}
		Pico-Cloud systems occupy a unique position in the distributed computing landscape. While inspired by conventional cloud and edge architectures, they diverge in technical assumptions and deployment philosophy. Table~\ref{table:comparison} summarizes these differences, which are further discussed below.
		
		\textbf{Scale and Performance.}  
		Public cloud platforms rely on massive, centralized data centers capable of dynamic scaling and delivering performance in the exaFLOP range. Edge computing clusters sit between the two extremes, often composed of compact but capable gateway devices or industrial PCs. Pico-Clouds remain at the micro scale, typically a handful of low-power nodes with aggregate performance measured in gigaflops. Their focus is not on raw throughput but on serviceability in constrained environments.
		
		\textbf{Hardware and Infrastructure Philosophy.}  
		Cloud providers build on server-grade hardware with redundancy and uniformity. Edge deployments generally use hardened COTS appliances engineered for reliability. Pico-Clouds instead rely on hobbyist-grade SBCs such as Raspberry Pi boards or low-cost microcontrollers. The trade-off is clear: performance and standardization are sacrificed in favor of affordability, accessibility, and adaptability to local conditions.
		
		\textbf{Connectivity and Dependence.}  
		Cloud services assume always-on, high-speed internet. Edge nodes typically require stable LAN or backhaul links but can buffer or function briefly offline. Pico-Clouds, by contrast, are explicitly designed to remain operational with intermittent or no internet access. This makes them particularly valuable in rural, mobile, or emergency scenarios where connectivity is unreliable or undesirable.
		
		\textbf{Operational Model and Control.}  
		Cloud deployments are centrally orchestrated and monitored, often with full automation and self-healing. Edge devices are commonly provisioned remotely through cloud-based control planes. Pico-Clouds emphasize user management: they are configured manually or with lightweight scripts, and control rests locally with the operator. This fosters autonomy and simplicity but comes at the expense of advanced orchestration features.
		
		\textbf{Security and Isolation.}  
		Cloud environments enforce strict multi-tenancy through hypervisors, IAM policies, and hardware-backed isolation. Edge systems may support limited multi-tenancy but are often scoped to a single organization. Pico-Clouds generally operate as single-tenant systems, relying on container-based isolation and process controls rather than hardware security. Although advanced features like secure boot are rare, direct physical control of devices is often stronger.
		
		\textbf{Use Case Focus.}  
		Cloud infrastructures target large-scale workloads such as SaaS delivery, machine learning training, or enterprise analytics. Edge systems are optimized for latency-sensitive services, content delivery, and real-time analytics near the data source. Pico-Clouds focus on narrower niches: lightweight, localized deployments for education, field operations, prototyping, and privacy-preserving personal services.
		
		Overall, Pico-Clouds complement rather than compete with traditional cloud and edge models. Their distinguishing feature is the ability to provide meaningful cloud-like functionality in contexts where resources, connectivity, or cost prohibit conventional infrastructure.

\section{Conclusion and Future Directions}
This paper presented the Pico-Cloud paradigm, a minimalist, edge-native cloud infrastructure constructed from ultra-compact, low-power computing platforms. By adapting cloud principles to resource-constrained environments, Pico-Clouds enable localized and energy-efficient service delivery in scenarios where centralized cloud or conventional edge systems are impractical.

The analysis of design choices, deployment models, and performance considerations indicates that Pico-Clouds can support diverse applications, including rural connectivity, disaster recovery, edge AI inference, and community-owned infrastructures. Although limited by processing capacity, storage endurance, and fault recovery, Pico-Clouds demonstrate practical strengths in modularity, affordability, and sustainability.

Future work should address several directions. Optimized orchestration frameworks for constrained hardware may enable finer scheduling and faster recovery without excessive overhead. Storage models tolerant to high latency and intermittent connectivity would improve reliability. Hardware–software co-design, including secure enclaves, energy-aware bootloaders, and passive thermal management, could extend robustness and system lifetime. Pico-Clouds also open research opportunities in federated learning, decentralized networks, and humanitarian deployments where operational constraints are critical.

Pico-Clouds extend the distributed computing spectrum by demonstrating that essential cloud-native capabilities—such as container orchestration, workload distribution, and local data processing—can be delivered effectively on ultra-low-power platforms. Rather than replacing cloud or edge infrastructures, they complement them by addressing scenarios where cost, portability, and connectivity constraints dominate system requirements.

\balance

\bibliographystyle{IEEEtran}
\bibliography{PicoCloud}

\end{document}